\documentclass[aps,superscriptaddress,amsmath,twocolumn,pra]{revtex4}
\usepackage[T1]{fontenc}
\usepackage[latin9]{inputenc}
\usepackage{amssymb}
\usepackage{graphicx}
\usepackage{subfigure}
\usepackage{epsfig}
\usepackage{txfonts}
\usepackage{amsfonts}
\usepackage{dcolumn}
\usepackage{esint}
\usepackage{mathrsfs}
\usepackage{bm}
\usepackage{amsmath}
\allowdisplaybreaks[1]

\begin{document}
\title[]{Searching for high-frequency gravitational waves by ground high field magnetic resonant sweepings}
\author{H. Zheng}
\address{School of Physics, State Key Laboratory of Optoelectronic Materials
and Technologies, Sun Yat-sen
University, Guangzhou 510275, China}

\author{L. F. Wei\footnote{E-mail: weilianfu@gmail.com}}
\address{School of Physics, State Key Laboratory of Optoelectronic Materials
and Technologies, Sun Yat-sen
University, Guangzhou 510275, China}
\address{Information Quantum Technology Laboratory, School of Information Science and Technology, Southwest Jiaotong University, Chengdu 610031, China}

\author{H. Wen}
\address{Institute of Gravitational Physics and Department of Physics, Chongqing University, Chongqing 400044, China}

\author{F. Y. Li}
\address{Institute of Gravitational Physics and Department of Physics, Chongqing University, Chongqing 400044, China}

\begin{abstract}
With laser interferometers, LIGO-Virgo collaboration has recently realized the direct detections of the intermediate-frequency (i.e., from dozens to hundreds of Hertz) gravitational waves (GWs) by probing their mechanically-tidal responses. Alternatively, in this letter we propose a feasible approach to actively search for the high-frequency GWs by probing their electromagnetic responses (EMRs) in a high alternating magnetic field.
 Differing from the original Gertsenshtein-Zeldovich configuration (in which the EMRs are proportional to the square of the amplitudes of the GWs, and consequently are too weak to be detected experimentally), the EMRs of the GWs passing through the present configuration are linearly related to the amplitudes of the GWs and thus the relevant signals are detectable with the current weak-light detection technique.
 As the wave impedances of the GWs-induced electromagnetic signals (EMSs) are very different from those the EM radiations in flat space-time, i.e, ($\cong 377 \Omega$), the stronger background noises (without any GWs information) could be effectively filtered out by using wave-matching technique.
 Given the frequency of the applied alternating magnetic field is conveniently adjustable, the configuration proposed here could be utilized to actively search for the GWs (if they really exist) in a sufficiently-wide frequency band (e.g., could be $10^7-10^{12}$Hz), once the scale of the cavity and the sweeping frequency of the applied alternating magnetic field are experimentally achievable.

PACS number(s): 03.65.Aa, 03.65.Ta, 03.67.Mn

\end{abstract}
\maketitle

{\it Introduction.--}It is well-known that, the existence of gravitational waves (GWs), i.e., the ripples of the curved space-time~\cite{book}, is one of the most important predictions in general relativity. Therefore, the direct detections of the GWs not only provided new evidences to verify the Einstein's gravitational theory, but also begun the era of gravitational wave astronomy. Yet, it is very difficult to verified such a prediction, due to the significantly-minute amplitudes of the GWs. Interestingly, with the precise laser interferometer the LIGO-Virgo collaboration~\cite{Gw150914, Gw151226, Gw170104, Gw170814} has realized recently the direct detections of the GWs by probing their minute mechanical displacements with the peak strains being at the order of $10^{-21}$ and the frequencies being in the low- and intermediate frequency band (i.e., dozens and hundreds of Hertz). Still, probing the GWs in wider bands, such as the high frequency ones expected by a series of models for cosmology and high-energy astrophysical processes~\cite{Giovannini a, Giovannini b, relic a, relic b}, is a challenge.

Historically, many approaches have been tried to directly probe the GWs or observe their certain indirect effects. In fact,
before the recent LIGO-Virgo's direct detections, an indirect experimental evidence of the existence of the GWs was obtained by observing the change in the orbital period of the PSR B1913+16 binary~\cite{PSR B 1913+16 a, PSR B 1913+16 b}, i.e., a pair of stars with one of them being a pulsar. Also, observing the $B$-mode polarizations in the cosmic microwave background (CBM) is regarded as another approach to indirect verify the existence of the GWs in very low frequency band~\cite{B-mode}.
Since 1980s the laser interferometers, including the typical LIGO-Virgo setup~\cite{LIGO, aLIGO, Virgo} and the proposed Laser Interferometer Space Antenna (LISA)~\cite{LISA1, LISA2}, and Einstein telescope~\cite{Einstein telescope 1,Einstein telescope 2, Einstein telescope 3}, etc. have been demonstrated to implement the detections of the GWs-induced mechanical displacements. Similarly, the pulsar timings and pulsar timing arrays are believed as another astronomic candidate to probe the GWs in low-frequency band~\cite{Detweiler1979}.

Besides the mechanically-tidal effects used in the above GW-detectors, it is believed that the electromagnetic responses (EMRs) of the GWs could be also observed. This is based on the so-called Gertsenshtein-Zeldovich (GZ) effects~\cite{G-Z effect}, i.e., the mutual conversions between EMWs and GWs in a high stationary magnetic field. Unfortunately, the amplitude of such a GW-induced EMW is too minute to be experimentally detected~\cite{too minute}, as the relevant detectable quantities are proportional to the square of the perturbative matric of the passing GWs~\cite{second-order 1, second-order 2}. To overcome such a difficulty, a series of modified schemes have been proposed by introducing certain ground auxiliary EMWs; the plane EMWs~\cite{auxiliary EMWs 1} and the Gaussian beams~\cite{auxiliary EMWs 2, auxiliary EMWs 3}, to enhance the observable effects of the perturbative EMW signals.

Continuously, in this letter we propose a feasible approach to actively search for the desired first-order EM perturbative signals generated by the high-frequency GWs passing through a high magnetic field. Besides a high stationary magnetic field utilized in the original GZ configuration, a relatively-weak alternating magnetic field is additionally applied. By exactly solving the relevant Einstein-Maxwell equation in a curved space-time, we show that the GWs-induced EMRs in the present setup are linearly related to the amplitudes of the passing GWs. These signals could be detected with the current weak-signal probing technique~\cite{d1,d2,d3,d4}. As the amplitude and the frequency of the applied weak alternating magnetic field are locally controllable, the scheme proposed here could be applied to actively search for the probably-existing high-frequency GWs (predicted by a series of theoretical models) in a sufficiently wide frequency band. Also, due to the wave impedances of the GWs-induced EMW signals, satisfying the Einstein-Maxwell equation in the curved space-time, are very different from those of the background EM noises (which obey the flat space-time Maxwell equation and thus are always take the value of $377 \Omega$ in vacuum), the well-developed wave impedance matching technique~\cite{Wave impendance 1, Wave impendance 2} could be utilized to safely filter out the background EM noises. As a consequence, only the GWs-induced EM signals, i.e., the signal photon fluxes, are left to be conducted into the detectors for detections.

{\it Model and solutions.---}\label{SEC:II}
A simplified configuration of our setup to actively search for the GWs is shown in Fig.~1,
\begin{figure}[t]
\begin{center}
\subfigure{\includegraphics[width=0.35\textwidth]{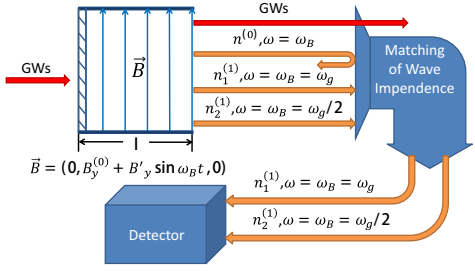}}
\end{center}
\vspace{-0.5cm}
\caption{A simplified configuration of a ground setup to search for the high frequency GWs along the $z$ axis: a single-side transmission cavity, i.e.,  region II $(0\leq z \leq l)$, is biased by an alternating magnetic field $B(t)$, and only the desired GWs-induced EM signals (with the frequencies $\omega_g=\omega_B, 2\omega_B$) can pass through the wave-impedance matchers for being detected.}
\end{figure}
wherein a background magnetic ${\bf B}(t)=(0, B^{(0)}_{y}+B'_{y}\sin(\omega_{B}t), 0)$ is confined in region II. $B^{(0)}_y$ is the high static magnetic field and $B'_y(t)=B'_{y}\sin\omega_{B}t$ with $B'_y(<B^{(0)}_y)$ being the alternating one of the frequency $\omega_B$. For the simplicity, the transmission of the cavity photons along the $-z$ direction is neglected. If a monochromatic circular polarized plane GW, i.e., a perturbed ripple:
\begin{eqnarray}
\left\{
\begin{array}{cc}
h_{\oplus}=h_{xx}=-h_{yy}=A_{\oplus}\Lambda_{\omega_g}(z,t),\\
\\
h_{\otimes}=h_{xy}=h_{yx}=i A_{\otimes}\Lambda_{\omega_g}(z,t),
\end{array}
\right.
\end{eqnarray}
with $\Lambda_{\omega_g}(z,t)=\exp[i\omega_g(z/c-t)]$, passes through the cavity, then the background flat Minkowski space-time will be perturbed as
\begin{gather}
g_{\alpha\beta}=\eta_{\alpha\beta}+h_{\alpha\beta}=\begin{bmatrix}
1\\
&-1+h_{\oplus}&h_{\otimes}\\
&h_{\otimes}&-1-h_{\oplus}\\
&&&-1
\end{bmatrix}.
\end{gather}
The electromagnetic field described by the tensor (2) obeys the relevant Einstein-Maxwell equation
\begin{eqnarray}
\left\{
\begin{array}{cc}
\partial_{\nu}\sqrt{{-g}} F^{\mu\nu}=\partial_{\nu}(\sqrt{{-g}}\ g^{\mu\alpha}g^{\nu\beta} F_{\alpha\beta})=0,\\
\\
\partial_{[\alpha}F_{\beta\gamma]}=0.
\end{array}
\right.
\end{eqnarray}
Above, as the perturbations of the ${\oplus}$- and ${\otimes}$ polarizations of the GW are individual, we just discuss the ${\oplus}$-polarization modulations specifically (the ${\otimes}$-polarization modulations can also be treated similarly).
It is well-known that, in a curved space-time only local measurements made by the observer traveling in the world-line are physical and all the observable quantities are just the projections of the relevant tensors on the tetrads of the observer's world-line. Obviously, in the present system the observer should be rest in the static magnetic field, i.e., only the zeroth component of
the four-velocity is non-vanish. Furthermore, with the simplified boundary conditions for the proposed configuration shown in Fig.~1, the solution of the Eq.~(3) can be conveniently divided into three parts; i.e.,
\begin{eqnarray}
\left\{
\begin{array}{ll}
&\widetilde{E}_{x}(z,t)=\widetilde{E}^{(0)}_{x}(z,t)+\widetilde{E}^{(1)}_{x}(z,t)
+\widetilde{E'}^{(1)}_{x}(z,t),
\\
&\widetilde{B}_{y}(z,t)=\widetilde{B}^{(0)}_{y}(z,t)+\widetilde{B}^{(1)}_{y}(z,t)
+\widetilde{B'}^{(1)}_{y}(z,t),
\end{array}
\right.
\end{eqnarray}
with the zero-order solution:
\begin{eqnarray}
\widetilde{E}^{(0)}_{x}(z,t)=-\dfrac{1}{2}B'_{y}c\sin[\omega_{B}(z/c-t)],\,
\widetilde{B}^{(0)}_{y}(z,t)=\frac{\widetilde{E}^{(0)}_{x}(z,t)}{c},
\end{eqnarray}
describing the usual electro-magnetic induction without the GWs perturbation.
The perturbative electromagnetic field, up to the first-order of the amplitude $A_{\oplus}$, can be expressed as:

(a) In regin I $(z<0)$:
\begin{eqnarray}
\widetilde{E}^{(1)}_{(x)}(z,t)=\widetilde{B}^{(1)}_{(y)}(z,t)=\widetilde{E}'^{(1)}_{(x)}(z,t)
=\widetilde{B}'^{(1)}_{(y)}(z,t)=0;
\end{eqnarray}

(b) in regin II $(0<z<l)$:
\begin{eqnarray}
\widetilde{E}^{(1)}_{(x)}(z,t)&=\dfrac{i}{2}A_{\oplus}B^{(0)}_{y}\omega_{g} z \Lambda_{\omega_{g}}(z,t),
\,\,
\widetilde{B}^{(1)}_{(y)}(z,t)&=\frac{\widetilde{E}^{(1)}_{(x)}}{c},
\end{eqnarray}
and
\begin{eqnarray}
\left\{
\begin{array}{cc}
\widetilde{E}'^{(1)}_{(x)}(z,t)=i\sum_{\alpha}\alpha\Phi_{\alpha}[\exp(-i\alpha\omega_{B}z)-1]
\Lambda_{\Delta_{\alpha}}(z,t),\\
\\
\widetilde{B}'^{(1)}_{(y)}(z,t)=i\sum_{\alpha}\alpha\Psi_{\alpha}[\exp(-i\alpha\omega_{B}z)-1]
\Lambda_{\Delta_{\alpha}}(z,t),
\end{array}
\right.
\end{eqnarray}
with $\Phi_{\alpha}=A_{\oplus}B'_{y}\omega_{g}\Delta_{\alpha} c/[2(\omega^{2}_{g}-\Delta^{2}_{\alpha})],\,
\Psi_{\alpha}=A_{\oplus}B'_{y}[\omega^{2}_{g}+\Delta^{2}_{\alpha}]/[4(\omega^{2}_{g}-\Delta^{2}_{\alpha})]$,
$\Delta_{\alpha}=\omega_g+\alpha\omega_B,\,\,\alpha=+,-$; and

(c) in regin III $(z>l)$:
\begin{eqnarray}
\widetilde{E}^{(1)}_{(x)}(z,t)=\dfrac{i}{2}A_{\oplus}B^{(0)}_{y}\omega_{g} l \Lambda_{\omega_{g}}(z,t),\,
\widetilde{B}^{(1)}_{(y)}(z,t)=\frac{\widetilde{E}^{(1)}_{(x)}}{c},
\end{eqnarray}
with $l=(2m+1)\pi c/\omega_{B}, m=0,1,2,...$ and
\begin{eqnarray}
\left\{
\begin{array}{cc}
\widetilde{E}'^{(1)}_{(x)}(z,t)=-2i\sum_{\alpha}\alpha \Phi_{\alpha}\Lambda_{\Delta_{\alpha}}(z,t),\\
\\
\widetilde{B}'^{(1)}_{(y)}(z,t)=-2i\sum_{\alpha}\alpha \Psi_{\alpha}\Lambda_{\Delta_{\alpha}}(z,t).
\end{array}
\right.
\end{eqnarray}

It is seen that, besides the usual electro-magnetic inductions in the local flat space-time, the perturbed electromagnetic field induced by the GWs passing through the alternating magnetic field includes three frequencies; $\omega_g$ and $\omega_g\pm\omega_B$. Distinguishing these signals from ones without any GW information, rather than the mechanical one utilized usually in the most GW detecting setups (such as the LIGO-Virgo installations), the GWs could be detected electromagnetically.

{\it Observable effects of GWs-induced perturbed electromagnetic signals.---}
Physically, the averaged power densities: $\langle S \rangle=T^{-1}\int^{T}_{0}S dt,\,\vec{S}=\vec{E}\times\vec{B}/\mu_0$, with $T$ being the period of the EMWs, could be used to describe the strengths of the electromagnetic signals. For a detector at the plane $z=z_0$ with the receiving surface $x=[x_1, x_2],\,y=[y_1, y_2]$, the averaged number of the photons with the frequency $\omega$ being detected per-second can be calculated as
\begin{equation}
n=\frac{1}{\hbar \omega}\int^{x_{2}}_{x_{1}}\int^{y_{2}}_{y_{1}}\langle S(z_0)\rangle dx dy.
\end{equation}
Here, $\hbar$ is Planck constant, and
%\begin{equation}
$S(z_0)=\widetilde{E}(z_0,t)\widetilde{B}_{y}(z_0,t)/\mu_0\simeq S^{(0)}(z_0)+S^{(1)}_1(z_0)+S^{(1)}_2(z_0)$.
%\end{equation}
The zero-order perturbative energy flow density, generated by the usual electro-magnetic induction of the alternating magnetic field, reads
%\begin{eqnarray}
$S^{(0)}(z_0)=\widetilde{E}^{(0)}_{x}(z_0)\widetilde{B}^{(0)}_{y}(z_0)/\mu_{0}$.
%\end{eqnarray}
The corresponding averaged photon number
\begin{equation}
n^{(0)}=\frac{\langle S^{(0)}\rangle_{\omega_B}}{\hbar\omega_{B}}=\frac{\Xi B'^{2}_{y}c}{8\mu_{0}\hbar \omega_{B}},\,\Xi=|x_2-x_1|\times|y_2-y_1|.
\end{equation}
is certainly very large. The energy flow densities of the first-order perturbative field (which is lineary relative to the amplitude of GWs) are
%\begin{eqnarray}
$S^{(1)}_1(z_0)=
[\widetilde{E}^{(0)}_{x}(z_0)\widetilde{B}^{(1)}_{(y)}(z_0)
+\widetilde{E}^{(1)}_{(x)}(z_0)\widetilde{B}^{(0)}_{y}(z_0)]/\mu_{0}$,
%\end{eqnarray}
and
%\begin{eqnarray}
$S^{(1)}_2(z_0)=
[\widetilde{E}^{(0)}_{x}(z_0)\widetilde{B}^{\prime(1)}_{(y)}(z_0)
+\widetilde{E}^{\prime(1)}_{x}(z_0)\widetilde{B}^{(0)}_{(y)}(z_0)]/\mu_{0}$,
%\end{eqnarray}
respectively. They correspond to two kinds of perturbed EM signals with the photon fluxes:
\begin{eqnarray}
n^{(1)}_{1}=\frac{\langle S^{(1)}_1(z_0)\rangle}{\hbar\omega_g}=\dfrac{\Xi B^{(0)}_{y}B'_{y}l}{4\mu_{0}\hbar }A_{\oplus},
\end{eqnarray}
generated by the perturbations of the GWs with the frequency $\omega_B=\omega_g$, and:
\begin{eqnarray}
n^{(1)}_{2}=\frac{\langle S^{(1)}_2(z_0)\rangle}{\hbar\omega_g}=\dfrac{3\Xi B'^{2}_{y}c}{4\mu_{0}\hbar\omega_{g}}A_{\oplus},
\end{eqnarray}
produced by the perturbations of the GWs with the frequency $\omega_g=2\omega_B$, respectively.
Specifically, Fig.~2 shows how the detectable number of the photons $n^{(1)}_1$ and $n^{(1)}_2$ versus the amplitude of the local alternating magnetic field $B'_y$ for the selected GWs signals.
\begin{figure}[htbp]
\begin{center}
\subfigure{\includegraphics[width=0.45\textwidth]{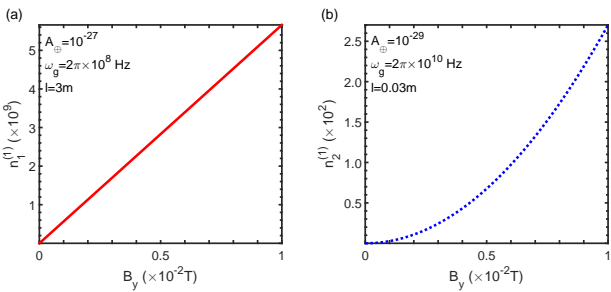}}
\setlength{\belowcaptionskip}{-1cm}
\end{center}
\vspace{-0.5cm}
\caption{The detectable photon number of the one-order perturbative signals $n^{(1)}_1$ and $n^{(1)}_2$ versus the amplitude $B'_{y}$ of the locally applied alternating magnetic field. The relevant parameters are chosen as: $B^{(0)}_{y}=10T$.}
\end{figure}
It is seen clearly that, without the alternating magnetic field (i.e., $B^{'}_y=0$) the first-order perturbative photon vanish (i.e., $n^{(1)}_1=n^{(1)}_2=0$). The lower frequency of the Gws, which corresponds to the higher amplitude of the GWs, and thus induces the stronger signal photon fluxes.
Additionally, the higher amplitude of the alternating magnetic field, applied locally to search for the GWs, yields the stronger electromagnetic responses of the GWs and consequently larger signal photon fluxes.
Anyway, with the single-photon detectors (see, e.g.,~\cite{d1,d2,d3,d4}) developed well in recent years, the induced photon fluxes could be detected, at least theoretically.

Next, with the usual frequency matching filtering technique, all the electromagnetic noises, except the ones with the frequency $\omega_B$, could be filtered out. However, the left signals with the same frequency $\omega_B$ still composite of three components; the photon fluxes $n^{(0)}$ without any GWs information, and the desired first-order perturbative ones, carrying the information of the GWs, $n^{(1)}_1$ and $n^{(1)}_2$. Fortunately, the wave impedances of the zero-order perturbative signals with the energy flow dengsity $S^{(0)}$ generated by the local electro-magnetic induction, satisfying the Maxwell equation in the flat space-time, always reads
\begin{equation}
Z_0=\frac{\mu_0 E^{(0)}_x}{B^{(0)}_y}=\sqrt{\frac{\mu_0}{\varepsilon_0}}\cong 377 \Omega.
\end{equation}
Here, $\varepsilon_0$ is the permittivity constant and $\mu_0$ the permeability constant in free space. While, the wave impedances of the first-order perturbed signals: $n^{(1)}_1$ and $n^{(1)}_2$, are
\begin{equation}
Z_1^{(1)}=\frac{\mu_0\widetilde{E}_x^{(0)}}{\widetilde{B}_x^{(1)}}=\frac{\mu_0B^{'}_yc^2}{B^{(0)}\omega_glA_{\oplus}},\,
Z_1^{'(1)}=\frac{\mu_0\widetilde{E}_x^{(1)}}{\widetilde{B}_x^{(0)}}=\frac{\mu_0 B^{(0)}\omega_glA_{\oplus}}{B^{'}_y},
\end{equation}
and
\begin{eqnarray}
Z^{(1)}_2=\dfrac{\mu_{0}\widetilde{E}^{(0)}_{x}}{\widetilde{B}'^{(1)}_{(y)}}
=\dfrac{6\mu_{0}c}{5A_{\oplus}},\,Z'^{(1)}_2=\dfrac{\mu_{0}\widetilde{E}'^{(1)}_{(x)}}{\widetilde{B}^{(0)}_{y}}
=\dfrac{2\mu_{0}A_{\oplus}c}{3},
\end{eqnarray}
respectively.
Fig.~3 shows how the wave impedances of the first-order perturbative electromagnetic signals versus the amplitude of the local alternating magnetic field $B'_y$. It is seen that $Z^{(1)}_1, Z^{(1)}_2\gg Z_0$ and $Z^{'(1)}_1, Z^{'(1)}_2\ll Z_0$. Therefore, with the wave impedance matching filtering technique, see, e.g.,~Refs.~\cite{Wave impendance 1, Wave impendance 2}, the electromagnetic signal $n^{(0)}$ without any GWs' information could be robustly filtering out, and consequently only the signals; $n^{(1)}_1$ and $n^{(1)}_2$, carrying the information of the GWs, could be effectively conducted into the weak light detectors to implement the desired detections.

\begin{figure}[htbp]
\begin{center}
\subfigure{\includegraphics[width=0.45  \textwidth]{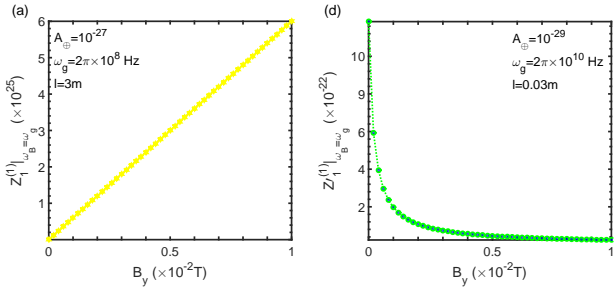}}
\end{center}
\vspace{-0.5cm}
\caption{The wave impedances $Z^{(1)}_{1}$ and $Z^{'(1)}_{1}$ of the first-order perturbative signals $n^{(1)}_1$ versus the amplitude of the applied alternating magnetic field $B^{'}_y$ in a high magnetic field $B^{(0)}_{y}=10T$ for typical frequenies and amplitudes of the GWs.}
\end{figure}

{\it Conclusions and Discussions.---}\label{SEC:IV}
In summary, a theoretical proposal to search for the high-frequency GWs with semi-open side cavity biased locally by an alternating high magnetic field is proposed. By analytically solving the Einstein-maxwell equation in the curved space-time, the electromagnetic responses of the GWs passing through the cavity are obtained. With the present setup the significantly stronger first-order perturbations of the GWs are generated and could be detected experimentally with the current developed-well weak-light detectors. Technically, the GWs-induced first-order perturbative EM signals could be selected out for detections, as the noises with the different frequencies and wave impedances could be filtering out by using the usual frequency matching and wave impedance matching filtering wave techniques.

It is emphasized that the proposed ground setup to search for the GWs works in a sufficiently wide band, as the frequency $\omega_B$ of the applied alternating magnetic field is adjustable within a very large regime, once the size of the cavity is experimentally achievable. Table I lists the typical frequency band of the detectable GWs with the proposed configuration. Here, a generic argument~\cite{Thorne}: $A_{\oplus}\propto\omega^{-1}$, on the relation between the amplitude and frequency of the GWs is used. One can see that the proposal should work for the GWs with the frequency from $10^7$Hz to $10^{12}$Hz, within the size of the cavity being experimentally reachable.

Certainly, the setup proposed here is still very imperfect for the realistic experiments. Its sensitivity should be further analyzed, in detail, in the presence of various noises and imperfect factors. Anyway, the present proposal provides an potential approach to locally deliver the observable first-order perturbative effects of the GWs passing through a high magnetic field. Probably, it can serve as an effective complementary of the very successful detections of the GWs based on their mechanically-tidal effects with the usual laser interferometers.

\begin{table}
\centering
\caption{Various parameters for probing the electromagnetic responses of GWs:
amplitude and frequency of the GWs, the size of the high magnetic field regime, and the number of signal photons at unit time. Here, $B'_{y}=0.005T$.}
\label{TAB1}
\begin{tabular}{c p{0.25cm} c p{0.25cm} c p{0.25cm} c p{0.25cm} c p{0.25cm} c p{0.25cm} c}
\hline\hline
  $A_{\oplus}$   && $\omega_{B} (Hz)$ &&      $l (m)$     && $n^{(1)}_{1}$         && $n^{(1)}_{2}$  \\
  \hline
  $10^{-26}$  && $2\pi\times10^{7}$   &&      $30$        &&$2.83\times10^{11}$   && $6.76\times10^{7}$  \\
  $10^{-27}$  && $2\pi\times10^{8}$   &&       $3$        && $2.83\times10^{9}$    && $6.76\times10^{5}$  \\
  $10^{-28}$  && $2\pi\times10^{9}$   &&      $0.3$       && $2.83\times10^{7}$    && $6.76\times10^{3}$  \\
  $10^{-29}$  && $2\pi\times10^{10}$  &&      $0.03$      && $2.83\times10^{5}$    && $67.6$              \\
  $10^{-30}$  && $2\pi\times10^{11}$  && $3\times10^{-3}$ && $2.83\times10^{3}$    && $0.68$              \\
  $10^{-31}$  && $2\pi\times10^{12}$  && $3\times10^{-4}$ && $28.3$                && $0.07$              \\
  \hline\hline
\end{tabular}
\end{table}

{\bf Acknowledgements.---} We thank Profs. Z. B. Li, X. G. Wu and Z. Y. Fang for useful discussions. This work is partly supported by the
National Natural Science Foundation of China (NSFC) under Grant Nos. U1330201.

\end{document}